# Electronic heat conductivity in a two-temperature state

Nikita Medvedev[1,2,*], Fedor Akhmetov[3], Igor Milov[4]

*1) Institute of Physics, Czech Academy of Sciences, Na Slovance 1999/2, 182 21 Prague 8, Czech Republic*

*2) Institute of Plasma Physics, Czech Academy of Sciences, Za Slovankou 3, 182 00 Prague 8, Czech Republic*

*3) Industrial Focus Group XUV Optics, MESA+ Institute for Nanotechnology, University of Twente, Drienerlolaan 5, 7522 NB Enschede, The Netherlands*

*4) ASML Netherlands B.V., Veldhoven, The Netherlands*

## Abstract

Laser irradiation of materials is most commonly modeled with the two-temperature model (TTM), or its combination with molecular dynamics, TTM-MD. For such modeling, the electronic transport coefficients are required. Here, we calculate the electronic heat conductivity at elevated electron temperatures up to 40,000 K. We apply the tight binding formalism to calculate the electron-phonon contribution to the electronic heat conductivity, and the linear response theory (in the single-pole Ritchie-Howie loss function approximation) for its electron-electron counterpart, implemented in the hybrid code XTANT-3. It allows us evaluation of the electronic heat conductivity in a wide range of materials – fcc metals: Al, Ca, Ni, Cu, Sr, Y, Zr, Rh, Pd, Ag, Ir, Pt, Au, and Pb; hcp metals: Mg, Sc, Ti, Co, Zn, Tc, Ru, Cd, Hf, Re, and Os; bcc metals: V, Cr, Fe, Nb, Mo, Ba, Ta, and W; other metals: Sn, Ga, In, Mn, Te, and Se; semimetal graphite; semiconductors – group IV: Si, Ge, and SiC; group III-V: AlAs, AlP, GaP, GaAs, and GaSb; oxides: ZnO, $TiO_2$, and $Cu_2O$; and others: $PbI_2$, ZnS, and $B_4C$.

## I. Introduction

Ultrashort laser pulse irradiation of materials drives the progress in fundamental and applied sciences [1–5]. The application of pulsed lasers enables elucidating basic phenomena in solid-state and soft-matter physics, nonequilibrium phenomena, and excited states of matter such as warm-dense matter state [6,7]. The major advantage of the femtosecond laser pulses is that the measurement time scales are comparable to the natural time window of the involved processes: the electron excitation and relaxation kinetics, electron-ion (electron-phonon) coupling, and transport effects [5,6,8].

From the technological viewpoint, laser irradiation is applied for materials processing, nano and micro technology, and medical applications such as laser surgery [4,5,9,10]. Given such a wide range of applications, it requires a detailed theoretical understanding and reliable models of the processes involved.

---

[*] Corresponding author: ORCID: 0000-0003-0491-1090; Email: nikita.medvedev@fzu.cz

Under ultrafast-laser irradiation, a sequence of effects takes place leading to observable material modifications. It starts with the photon absorption, which drives the electronic system of the target to transient nonequilibrium [2]. Excited electrons scatter among themselves, relaxing their distribution function to the equilibrium (Fermi-Dirac) one. This nonequilibrium stage is short-lived, and the electronic ensemble typically thermalizes at femtosecond timescales. Then, the energy exchange with the atomic system (often described in terms of phonons) takes place at picosecond timescales. During that time, a multitude of effects manifests itself, from possible nonthermal modification of the interatomic potential (typically, in covalent materials), to electronic and energy transport, bringing energy out of the laser spot. That stage may be characterized as the two-temperature state, in which the electronic system possesses a temperature high above the atomic one.

To describe this state and the processes involved, the most widely used model is the so-called two-temperature model (TTM), consisting of the two coupled energy transport equations [12,20]:

$$\begin{cases} C_e(T_e)\frac{\partial T_e}{\partial t} = \frac{\partial}{\partial x}\left(k(T_e,T_a)\frac{\partial T_e}{\partial x}\right) - G(T_e,T_a)(T_e - T_a) + S(t,x), \\ \qquad C_a\frac{\partial T_a}{\partial t} = \frac{\partial}{\partial x}\left(k_a\frac{\partial T_e}{\partial x}\right) + G(T_e,T_a)(T_e - T_a) \end{cases} \quad (1)$$

here $T_e$ is the electronic temperature, and $T_a$ is the atomic one; $C_e(T_e)$ is the volumetric electron heat capacity dependent on the electronic temperature (and only mildly dependent on the atomic temperature below the melting point [11]); $k(T_e,T_a)$ is the electron thermal conductivity, and $k_a$ is the atomic one; $G(T_e,T_a)$ is the electron-ion coupling; $S(t,x)$ is an external heat source (the laser pulse) [19]; and $C_a$ is the volumetric atomic heat capacity.

Nowadays, the gold standard in the field of ultrafast material science is a description of the atomic system with the molecular dynamics (MD) simulation instead of the simple thermodynamic equation (1). The MD technique enables tracing the nonequilibrium response of the atomic system and possible phase transitions, thereby making a major improvement upon the TTM [12,13].

Coupling the thermodynamical equation for a description of the transient state of the electronic system with the MD of the atomic one allows us to realistically trace the material response. Such models are known as coupled TTM-MD [1,14]. In such models, the energy provided by electrons is distributed to atoms *via* rescaling their velocities by a proportionality factor $\xi$ [13]:

$$M_i\frac{d^2\mathbf{R}_i}{dt^2} = -\frac{\partial V(\{R_{ij}\})}{\partial \mathbf{R}_i} + \xi M_i \mathbf{v}_i,$$

$$\xi = \frac{G}{N_{at}}\frac{(T_e - T_a)V_0}{\sum_j M_j v_j^2} \quad (2)$$

here $\boldsymbol{v}_i$ is the velocity of $i^{\text{th}}$ atom with mass $M_i$, $V_0$ is the volume of the MD simulation box, and $N_{at}$ is the number of atoms in the simulation box. These Eqs. (2) replace the second equation in set (1) describing the atomic system [1,13].

Both TTM and TTM-MD models require reliable parameters describing the two-temperature state of a material. The electronic heat capacity is straightforward to calculate knowing only the band structure of the material [15,16]. The electron-phonon coupling was previously reported for a wide range of materials: e.g., metals in Refs.[11,17,18]; and semiconductors in Ref.[19]. The last missing parameter of the TTM is the electronic heat conductivity. This problem forms the subject of the present study.

## II. Model

We use the XTANT-3 hybrid code for the calculation of the electronic thermal conductivity [20,21]. The code combines a few approaches: transferable tight-binding (TB) molecular dynamics, Monte-Carlo (MC) method, Boltzmann collision integrals, and Kubo-Greenwood formalism or the random-phase approximation (RPA) for evaluation of the optical properties [21,22]. This combined formalism allows us to obtain all necessary parameters for the evaluation of the electronic transport coefficients. The electron-phonon coupling parameter was reported previously in Refs.[17–19]. Here, we extend the method to evaluate the electronic heat conductivity dependent on the electron temperature.

The total electronic heat conductivity consists of two terms: the electron-ion (or electron-phonon, $\kappa_{e-ph}$) and the electron-electron ($\kappa_{e-e}$) scattering contributions combined using the Matthiessen's rule [23]:

$$\kappa_{tot} = \left(\frac{1}{\kappa_{e-ph}} + \frac{1}{\kappa_{e-e}}\right)^{-1} \quad (3)$$

The electron-phonon part of the electronic heat conductivity is dependent on the electronic temperature ($T_e$) and may be calculated within the Kubo-Greenwood formalism [24]:

$$\kappa_{e-ph} = \frac{1}{T_e}\left(L_{22} - T_e \frac{L_{12}^2}{L_{11}}\right) \quad (4)$$

where the Onsager coefficients are [24]:

$$L_{ij} = -\frac{(-1)^{i+j}}{\Omega m_e^2} \sum \frac{df}{dE_k}(E_k - \mu)^{i+j-2} |\langle k|p|k'\rangle|^2 \quad (5)$$

Here $\Omega$ is the volume of the simulation box (supercell); $m_e$ is free electron mass; $E_k$ is the electronic energy level (eigenstate of the TB Hamiltonian); $f$ is the electronic distribution function (fractional electron populations of the energy levels $E_k$, which is assumed to be the Fermi-Dirac distribution, normalized to 2 due to spin degeneracy); $\mu$ is the electronic-temperature dependent chemical potential.

The momentum matrix elements in the Onsager coefficients are calculated the same way as previously used for the calculation of the optical coefficients in the RPA [25]:

$$|\langle n'|p|n\rangle|^2 = \left| \sum_{R,R',\sigma,\sigma'} B_{\sigma n}(R) P(R,R') B_{\sigma' n'}(R') \right|^2 \tag{6}$$

where $R$ are the coordinates of atoms; $\sigma$ are the atomic orbitals; and $B_{\sigma n}(R)$ and $B_{\sigma' n'}(R')$ are the corresponding eigenvectors of the TB Hamiltonian matrix $H(R, R')$. Using nonorthogonal Hamiltonian and overlap matrix, the momentum operator may be approximately obtained as [26]:

$$P(R, R') = \frac{m_e}{\hbar} \left( \frac{\partial H(R, R')}{\partial k} - E_n \frac{\partial S(R, R')}{\partial k} \right) \tag{7}$$

In the case of orthogonal Hamiltonian (as used for some materials), $S_{ij} = \delta_{ij}$. Using the operator identity, the momentum operator can be calculated within the TB as [25,26]:

$$P(R, R') = i \frac{m_e}{\hbar} [R - R'] (H(R, R') - E_n S(R, R')) \tag{8}$$

In this representation, only the atomic displacements are used for evaluation of the momentum matrix elements, thus, this term only describes the electron-atom heat conductivity.

The electron-electron scattering probabilities are not accessible within the TB formalism, but its contribution to the conductivity can be evaluated with the electron-electron scattering cross-section. We use the same formalism, Eq.(4), for the conductivity, with the Onsager coefficients in the following representation:

$$L_{ij}^{e-e} = -\frac{(-1)^{i+j}}{\Omega m_e^2} \sum \frac{df}{dE_k} (E_k - \mu)^{i+j-2} \frac{v\lambda}{3} \tag{9}$$

For the electron-electron scattering, the electron velocity, $v$, may be approximated as a free-particle velocity (here counted from the bottom of the valence/conduction band, consistent with the MC method employed in XTANT-3), and the inelastic mean free path, $\lambda$, can be obtained with the cross-section of electron-electron scattering. The electronic energy levels (and the corresponding chemical potential) from the TB calculations are used, the same as for the electron-phonon contribution evaluation above.

$$\lambda^{-1} = \sigma n_{at} = n_{at} \int_{W_-}^{W_+} \int_{Q_-}^{Q_+} \frac{d^2\sigma}{dWdQ} dWdQ,$$

$$\frac{d^2\sigma}{d(\hbar\omega)d(\hbar q)} = \frac{2(Z_e e)^2}{n_{at}\pi\hbar^2 v^2} \frac{1}{\hbar q} \left(1 - e^{-\frac{\hbar\omega}{k_B T_e}}\right)^{-1} Im\left[\frac{-1}{\varepsilon(\omega, q)}\right] \tag{10}$$

Where $\sigma$ is the electron-electron scattering cross section; $e$ is the electron charge; $k_B$ is the Boltzmann constant; $Z_e$ is the effective charge of the incident particle (for an electron, $Z_e = 1$); $n_{at}$ is the atomic density; $W=\hbar\omega$ is the transferred energy and $\hbar$ is Planck's constant; and the recoil energy in the non-relativistic limit is $Q = \hbar^2 q^2/(2m_t)$, with $\hbar q$ being the transferred momentum, and $m_t$ being the scattering center (electron) mass. The cross-section integration limits are [27,28]:

$$Q_\pm = \frac{m_{in}}{m_t}\left(\sqrt{E} \pm \sqrt{E-W}\right)^2,$$

$$\begin{cases} W_- = I_p \\ W_+ = \frac{4m_{in}m_t}{(m_{in}+m_t)^2} E \end{cases} \quad (11)$$

with $m_{in}$ being the mass of the incident particle (here, an electron, $m_{in} = m_e$); $I_p$ is the ionization potential of the atomic shell an electron is being ionized from ($I_p$=0 for the conduction or valence band of a metal/semiconductor; for core shells, atomic ionization potentials are used from EPICS2017 database [29]).

We use the same cross-sections as in the MC module of XTANT-3, namely, the complex dielectric function (CDF, $\varepsilon(\omega,q)$) (or linear response theory) within the Ritchie-Howie approach [30]. The imaginary part of the inverse CDF function (the loss function) entering Eq.(10) is approximated with the set of artificial oscillators according to the Ritchie-Howie formalism [27,30]:

$$Im\left(\frac{-1}{\varepsilon(\omega,0)}\right) \approx \sum_i \frac{A_i\gamma_i\hbar\omega_i}{((\hbar\omega_i)^2 - E_i^2)^2 + (\gamma_i\hbar\omega_i)^2} \quad (12)$$

In the single-pole approximation, only one oscillator is used for each atomic shell and the valence/conduction band [31]. The oscillator coefficients may be obtained according to the position of a collective mode of the particles oscillations (the plasmon mode for scattering on the valence/conduction band electronic system; ionization potential for core shells, $I_p$) [31]:

$$E_{0_{sp}} = \max(\hbar\Omega_p, I_p)$$

$$\gamma_{sp} = E_{0_{sp}} \quad (13)$$

$$A_{sp} = (\hbar\Omega_p)^2 / \int_0^\infty W Im\left(\frac{-1}{\varepsilon(W, Q=0, A=1)}\right) dW$$

Here, $\Omega_p^2 = 4\pi e^2 n_e/m_e$, with $n_e$ being the valence/conduction band electron density (or core-shell electron density). The $A_{sp}$, the normalization coefficient, is defined by the $k$-sum rule [31].

For the calculations, we use averaging over 7x7x7 $k$-points on the Monkhorst-Pack reciprocal space grid [32]. It seems to be sufficient for the convergence of the results for the simulation boxes used, typically containing a few hundred atoms (the number of atoms differs for different materials, as specified below). However, we must note that, similar to our previous calculations of the electron-phonon coupling parameter [17], the methodology is primarily developed for high electronic temperatures and is not expected to precisely reproduce the room-temperature values of such properties as thermal conductivity, which were not specifically used in fitting the TB parameters. We expect that with dedicated adjustment of the TB parameters, it should be possible to reproduce the room temperature values, but this is beyond the scope of the present work.

In the literature, often the Drude model is used as the standard approximation for the electronic thermal conductivity [1]:

$$k_{tot} \approx B k_0 \frac{T_e}{A T_e^2 + B T_a} \quad (14)$$

here the coefficients *A* and *B* are fitted to room-temperature experimental data [33]. A low-temperature approximation of it is a linear dependence, $k_{tot} \approx k_0 T_e / T_a$, that is also often used in the literature (with $k_0$ being the room-temperature conductivity). Below, we will use these approximations on several occasions for comparison with our results.

## III. Results

### 1. Analysis of the model

Let us start with the analysis of the contribution of the electron-phonon ($\kappa_{e-ph}$) and electron-electron ($\kappa_{e-e}$) contributions to the total electronic heat conductivity. We use 256 atoms in the supercell (4x4x4 orthogonal unit cell). For 128 atoms, the results were close (not shown), indicating convergence of the calculations with respect to the number of atoms in the simulation box.

In the shown examples of aluminum, gold, silicon, and ruthenium (Figure 2), the electron-phonon scattering contribution dominates at low electronic temperatures, whereas the electron-electron contributes mainly at high electronic temperatures (the same situation is observed in all other materials; not shown). Such a situation is expected, see e.g. Ref.[23], which validates our calculations.

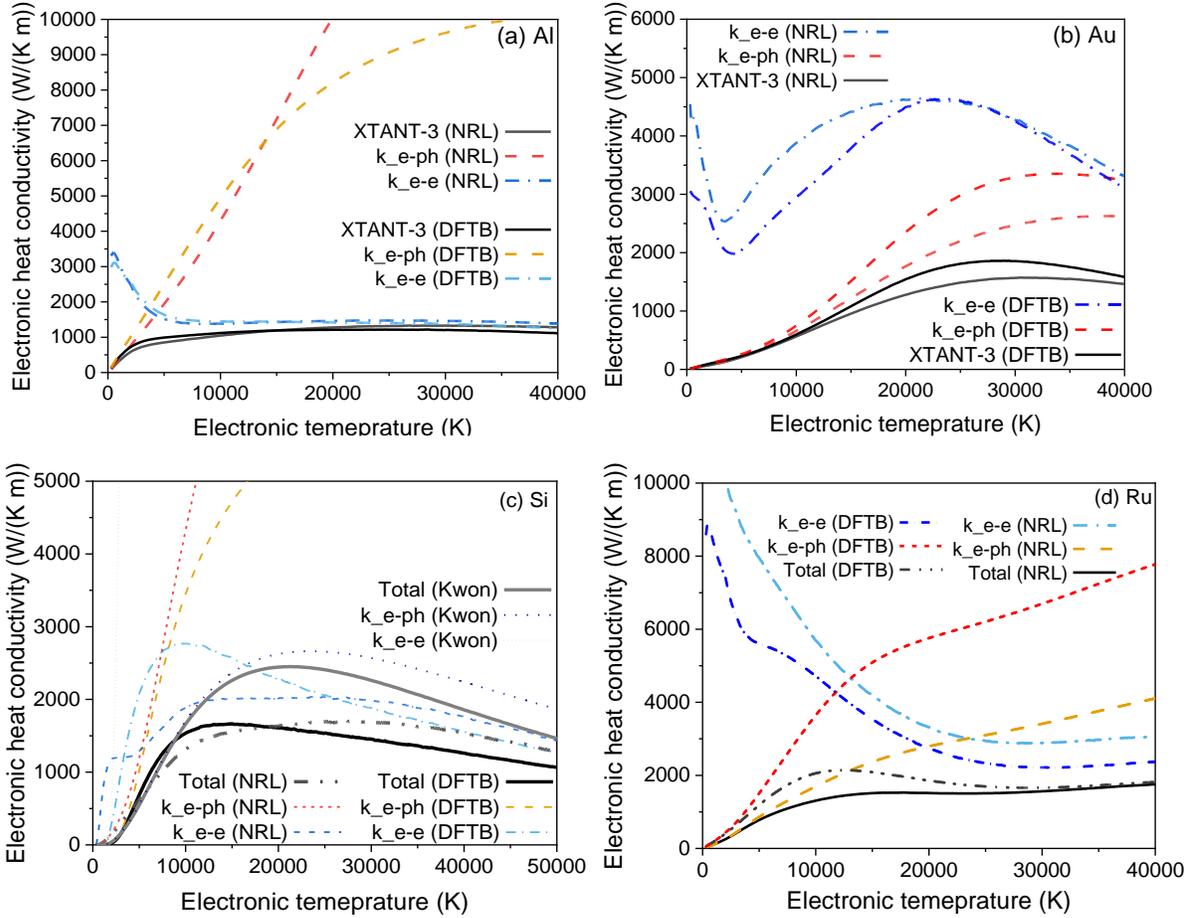

Figure 1 (color online). Electronic heat conductivity as a function of the electronic temperature in (a) aluminum, (b) gold, (c) silicon, and (d) ruthenium. The following TB parameterizations are compared: NRL[34] for all cases vs. DFTB matsci-0-3 for aluminum and silicon[35], auorg-1-1 for gold[36], and the one from Ref.[37] for ruthenium; additionally, orthogonal parameterization from Kwon *et al.* is shown for Si [38]. The total one and contributions of electron-phonon and electron-electron conductivities are shown.

Additionally, we analyze the influence of a particular tight binding parameterization on electronic heat conductivity. To that end, we compare the two types of $sp^3d^5$-based transferable TB parameterizations: NRL from Ref. [34] (which was used in the previous calculations of the electron-phonon coupling in metals [17]) and DFTB (matsci-0-3 for aluminum and silicon [35], auorg-1-1 for gold [36], one from Ref. [37] for ruthenium; and additionally an orthogonal $sp^3$-based parameterization from Kwon *et al.* is used for Si [38].), as also shown in Figure 2. The comparison of the DFTB and NRL TB parameterizations demonstrates very close results for all materials, suggesting that the sensitivity of the calculations to the particular TB parameterizations is relatively small. Kwon *et al.*'s parameterization for silicon produces a large deviation, which is a result of a smaller basis set used in it and thus a poorer band structure. The differences in the electronic heat conductivity are much smaller than those reported for the electron-phonon coupling calculations [11]. We proceed with the evaluation of the electronic heat conductivity for various materials.

## 2. Fcc metals

The following fcc metals are studied here: Al, Ca, Ni, Cu, Sr, Y, Zr, Rh, Pd, Ag, Ir, Pt, Au, and Pb (listed in the order of increasing atomic mass). For all the materials in this section, we used 256 atoms in the simulation box with NRL TB parameterization, unless specified otherwise [34]. For comparison with data available in the literature, we selected only those data that contained specifically electronic heat conductivity. The data on total heat conductivity (atomic and electronic) were not included if the two contributions could not be decoupled.

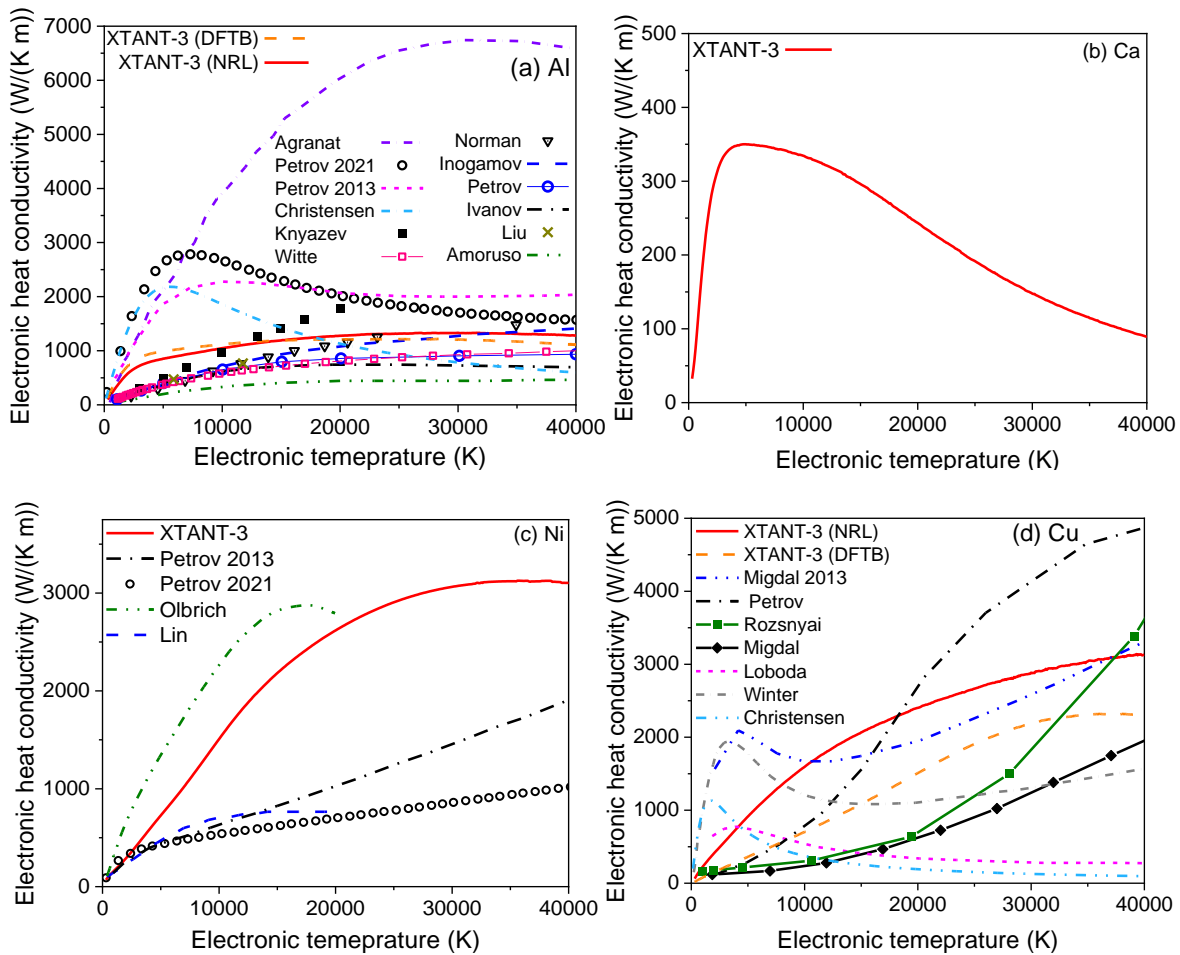

Figure 2. Electronic heat conductivity as a function of the electronic temperature. (a) In Aluminum, XTANT-3 calculated conductivity (with NRL and DFTB) vs. those from Petrov *et al.* 2021 [39]; Agranat *et al.* [40], Norman *et al.* [41], Inogamov and Petrov [42]; Ivanov and Zhigilei [13]; Amoruso *et al.* [43]; Petrov *et al.* 2013[44]; Petrov *et al.* 2018 [45]; Christensen *et al.*[46] (using Eq.(14)); Witte *et al.* [47]; Liu

*et al.* [48]; and Knyazev and Levashov (evaluated at the atomic temperature of 3000 K) [49]. (b) in Calcium. (c) in Nickel, XTANT-3 calculations compared to those from Lin and Zhigilei[50], Olbrich *et al*.[51]; Petrov 2021 *et al*.[39], and Petrov *et al.* 2013 [44]. (d) in Copper, XTANT-3 calculated conductivity compared with those by Migdal *et al*. 2013 [52]; Petrov and Davidson [53]; Rozsnyai [54]; Winter *et al*. [55]; Christensen *et al*.[46] (using Eq.(14)); Migdal et al. [56], and that constructed in[56] with parameters from Loboda et al.[57].

Figure 2a shows a comparison of the electron heat conductivity in aluminum calculated with XTANT-3 vs. those calculated by other authors found in the literature. The comparison shows that the initial growth of the XTANT-3-calculated conductivity is close to semi-empirical modeling by Petrov *et al.* (2021) [39], Agranat *et al.* [40], and Christensen *et al.*[46] (using Eq.(14)); whereas at high electronic temperature, it is closer to the results by Norman *et al.* [41] (which employed Kubo-Greenwood ab-initio-based calculations) and Inogamov and Petrov [42] (which used a semiempirical Boltzmann-based model). In the absence of experimental data on the electronic heat conductivity at high electronic temperatures in a two-temperature state, it is not possible to evaluate which model performs better.

Figure 2b shows the electronic heat conductivity in calcium. Calcium shows much lower electronic heat conductivity even at its peak, decreasing further due to the electron-electron interaction at the studied temperatures. Unfortunately, we found no data to compare it with.

Figure 2c demonstrates the XTANT-3 calculated electronic heat conductivity in nickel. At low electronic temperatures, all the calculations agree. With the increase of $T_e$, the XTANT-3 predicted curve rises closest to the modeled data by Olbrich *et al*.[51].

Calculated electronic heat conductivity in copper is shown in Figure 2d. There is a noticeable difference between the XTANT-3 calculations with the NRL parameterization and the DFTB matsci-0-3. At low electronic temperatures, the DFBT-based XTANT-3 calculations are best matched with the most recent average-atom calculations by Petrov and Davidson [53]. Overall, the data found in the literature exhibit wide discrepancies, including qualitatively different behavior, which indicates that experimental validation is urgently needed.

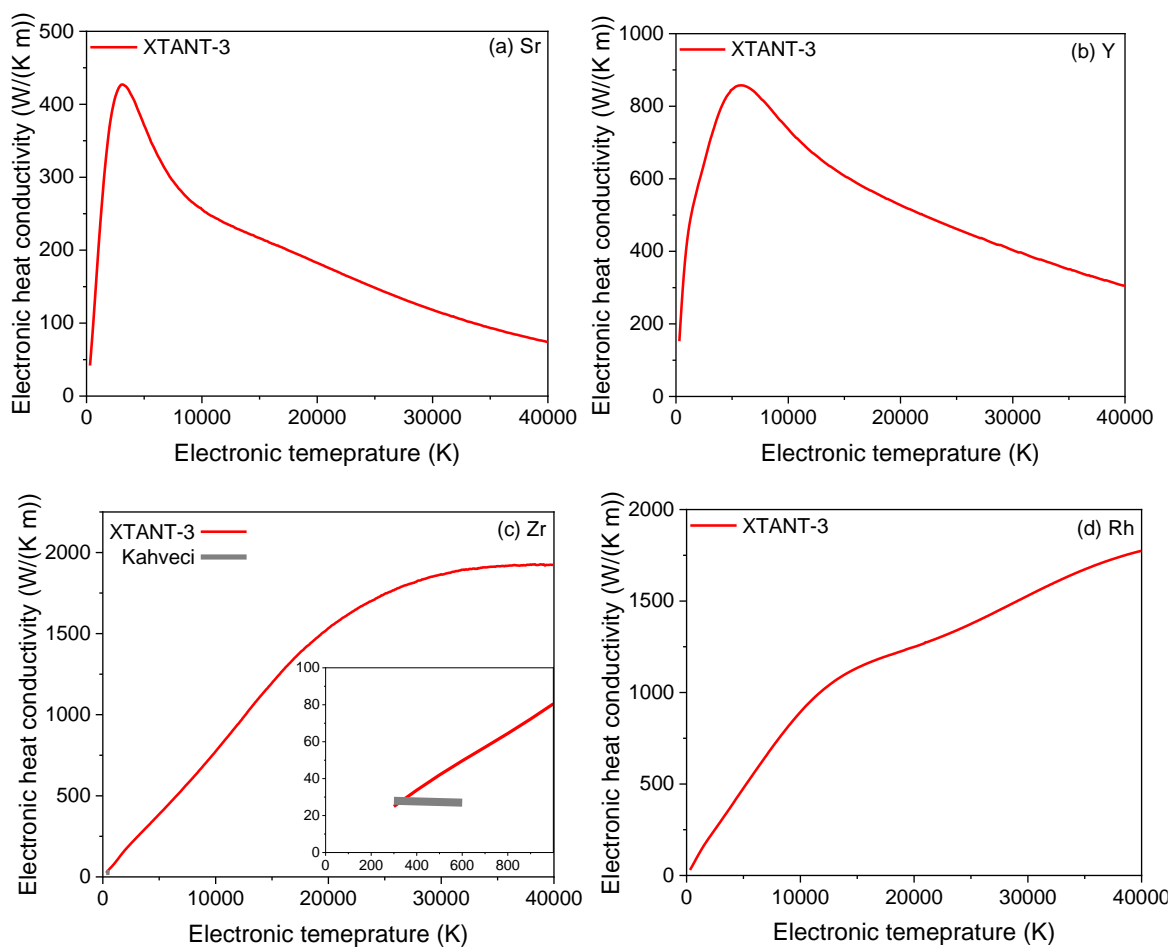

Figure 3. Electronic heat conductivity as a function of the electronic temperature, calculated with XTANT-3. (a) In Strontium, (b) in Yttrium. (c) in Zirconium, compared to near-room temperature estimate by Kahveci *et al*.[58] (inset zooms onto the low-temperature region). (d) in Rhodium.

Figure 3 shows the XTANT-3 calculated electronic heat conductivity in strontium, yttrium, zirconium, and rhodium. Only low-temperature data for Zr were found in the literature. The XTANT-3 calculations agree well at the near room temperature with the data from[58] (see inset in Figure 3c), however, showing an immediate increase, whereas Kahveci *et al*. predicted a nearly constant value in the region from the room temperature up to 600 K.

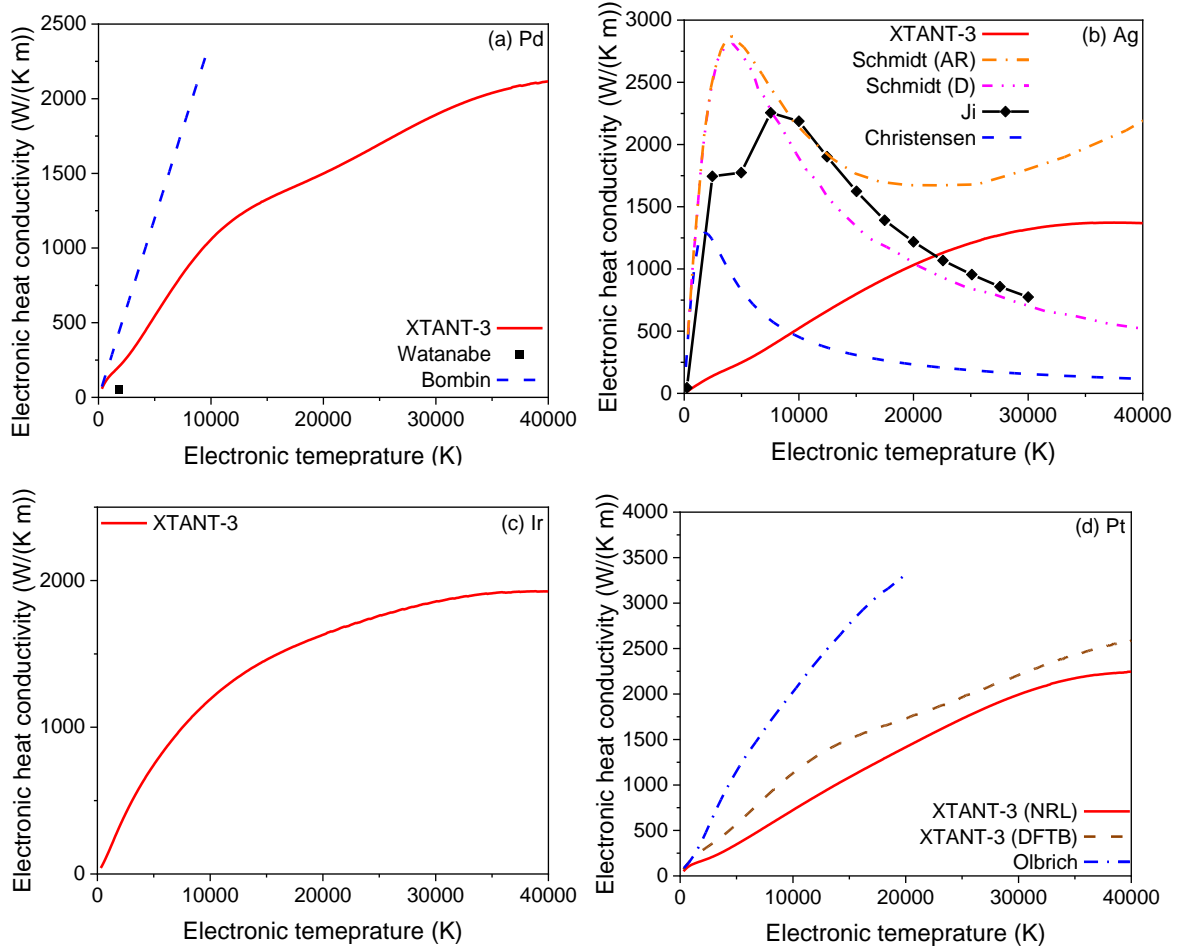

Figure 4. Electronic heat conductivity as a function of the electronic temperature. (a) In Palladium, XTANT-3 calculated conductivity vs. the linear approximation from[59]; and the data on Pd melt at 2000 K by Watanabe *et al.*[60]. (b) in Silver, XTANT-3 data compared with calculations by Ji and Zhang [61]; Christensen *et al.* [46] and Schmidt *et al.* (marked 'D' for calculation with Drude model using Eq.(4)(c), and 'AR' from Anisimov and Rethfeld using Eq.(4)(d) in Ref. [62]). (c) in Iridium. (d) in Platinum, calculated with XTANT-3 using NRL and DFTB parameterization from Ref.[37], compared to the modeled data from Olbrich *et al*. [51].

Figure 4 presents the calculation results for palladium, silver, iridium, and platinum. For palladium, the XTANT-3 predicted electronic thermal conductivity is compared with the linear approximation from Ref.[59], and with the data on palladium melt at 2000 K by Watanabe *et al.*[60]. In the melted state, the heat conductivity is expected to be lower than that in the solid state at room atomic temperature (which is used in the calculations). The linear approximation agrees with our calculations at near-room temperature, however, the deviation from the linear dependence (an approximation to Eq.(14)) used in Ref.[59] starts very early.

Silver data in Figure 4b are compared with those from Ji and Zhang [61], who used the Drude model for the evaluation of the electronic heat conductivity, thus a large difference with our method can be expected. Another comparison is with Eq.(14) using the coefficients from Ref.[46].

This expression predicts a much faster increase and then decrease of the electronic thermal conductivity in comparison with XTANT-3 calculations in silver.

For platinum in Figure 4d, two TB parameterizations are compared, the NRL [34] and DFTB from Ref.[37]. They are reasonably close to each other. A comparison with the model by Olbrich *et al.*[51] shows both XTANT-3 calculations predict a somewhat slower increase of the electronic heat conductivity. For this comparison, the thermal diffusivity coefficient presented in Ref. [51] was converted to the heat conductivity *via* the temperature-dependent electronic heat capacity calculated with XTANT-3.

Figure 5a shows calculated electronic heat conductivity in gold. Comparison with other models shows a behavior similar to the high-temperature calculations by Norman *et al.* performed with the Kubo-Greenwood formalism [41]. The other calculations, based on models similar to Eq.(14), predict a peak at low electronic temperatures, absent in Kubo-Greenwood calculations. Again, similar to the cases above, the differences between the existing calculations are very large. The results by Karna *et al.*[63] reach unrealistically high values of $K\sim 90,000$ W/(Km) at $T_e\sim 35,000$ K (not shown). We also note that XTANT-3 calculations do not reproduce the room temperature value well in this case.

Figure 5b presents the calculated electronic heat conductivity in lead. In contrast to gold, here the room temperature value is reproduced reasonably well (49 W/(m K) vs. 35 W/(m K) in experiments).

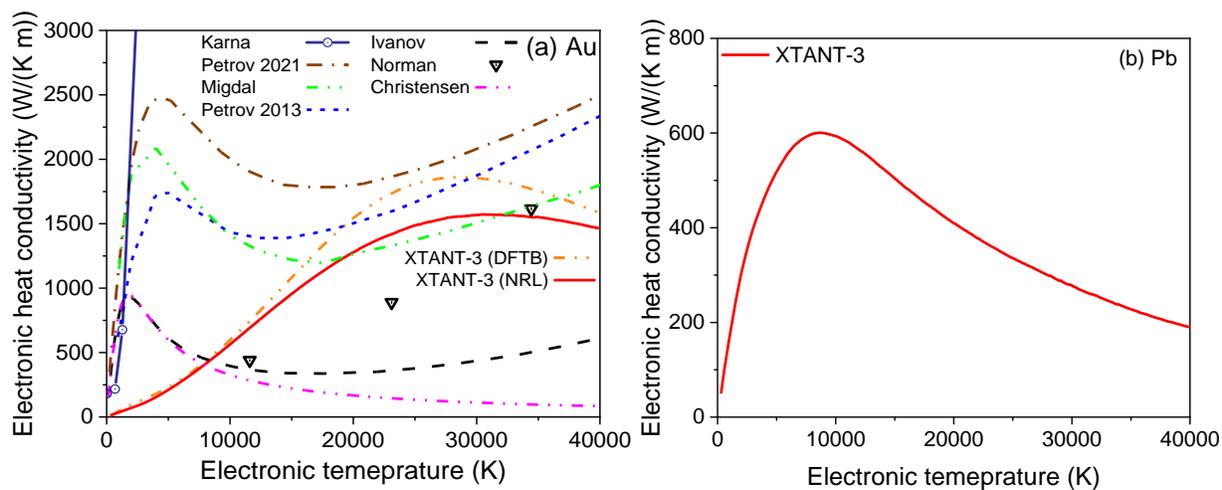

Figure 5. Electronic heat conductivity as a function of the electronic temperature. (a) in Gold, the XTANT-3 calculated conductivity vs. those by Norman *et al.*[41]; Ivanov and Zhigilei [13]; Petrov *et al.* 2021[64]; Petrov *et al.* 2013 [44]; Migdal *et al.*[52]; Christensen *et al.*[46] (using Eq.(14)); Karna *et al.*[63] (these results reach $K\sim 90,000$ W/(Km) at $T_e\sim 35,000$ K, and thus are cut off in the plot). (b) in Lead, calculated with XTANT-3.

## 3. hcp metals

The following hcp metals are analyzed here: Mg, Sc, Ti, Co, Zn, Tc, Ru, Cd, Hf, Re, and Os. In all cases, the supercell contained 256 atoms (4x4x4 orthogonal unit cells). NRL tight binding parameterization is used [34]; for ruthenium, additionally, a DFTB one from Ref.[37] is studied for comparison.

Figure 6 shows the calculated electronic heat conductivity in magnesium, scandium, titanium, and cobalt. For titanium, the calculation results are compared with the available data on Ti melt at 2000 K by Watanabe *et al.*[60]. Similar to the case of palladium discussed above, the heat conductivity in the melt is expected to be below the calculated in a solid state at room atomic temperature, and thus such a comparison looks reasonable. There are no data available for comparison for the other materials in Figure 6.

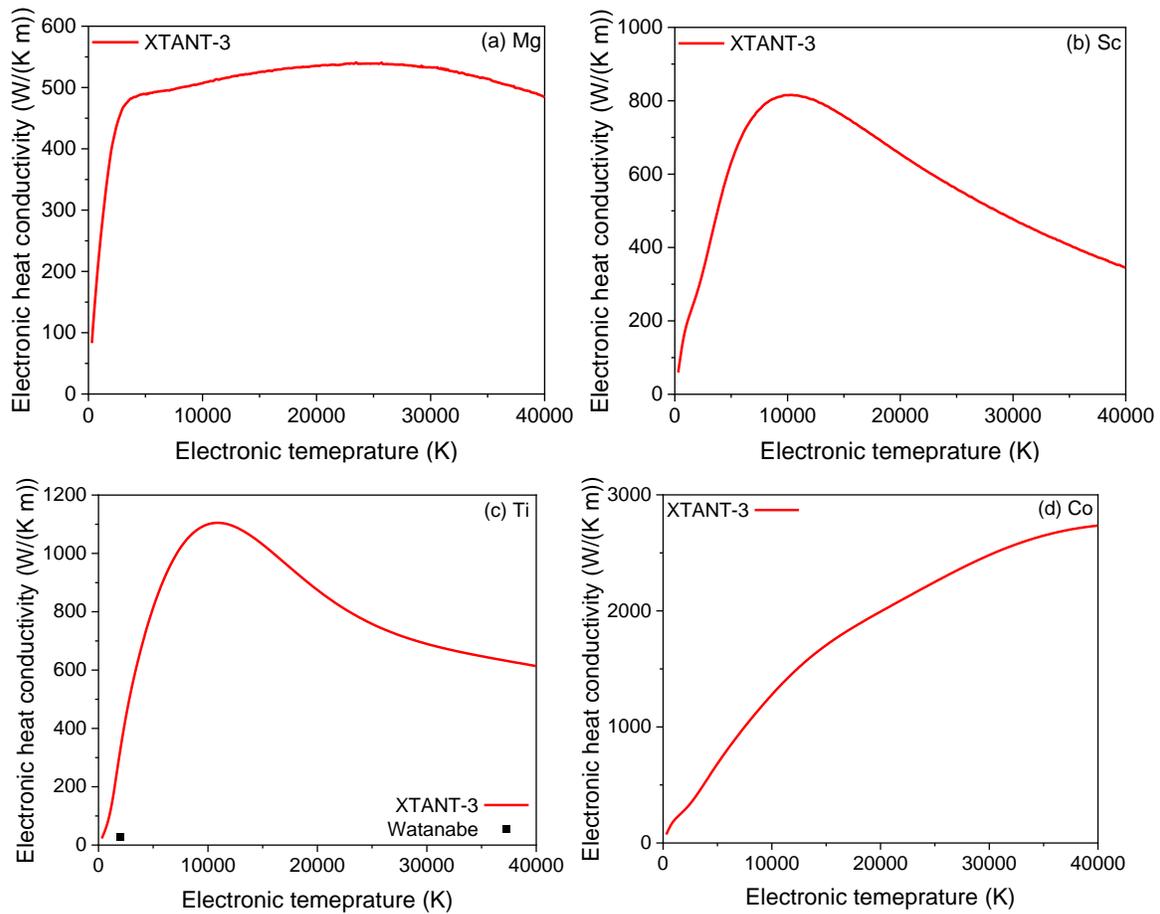

Figure 6. Electronic heat conductivity as a function of the electronic temperature. (a) in Magnesium. (b) in Scandium. (c) In titanium, XTANT-3 calculations are compared with the data on Ti melt at 2000 K by Watanabe *et al.*[60]. (d) in Cobalt.

Figure 7 shows the calculated electronic heat conductivity in zinc, technetium, ruthenium, and cadmium. In ruthenium, a comparison with other models shows a very large discrepancy, however, we must note that the two other curves were obtained by fitting the model parameters (Eq.(14) by Milov *et al.*[65]) to the experimental data on the equilibrium case ($T_a=T_e$), whereas XTANT-3 calculations are performed in two-temperature state ($T_a>T_e$), thus the differences are to be expected. No data for comparison were found for other materials shown in Figure 7.

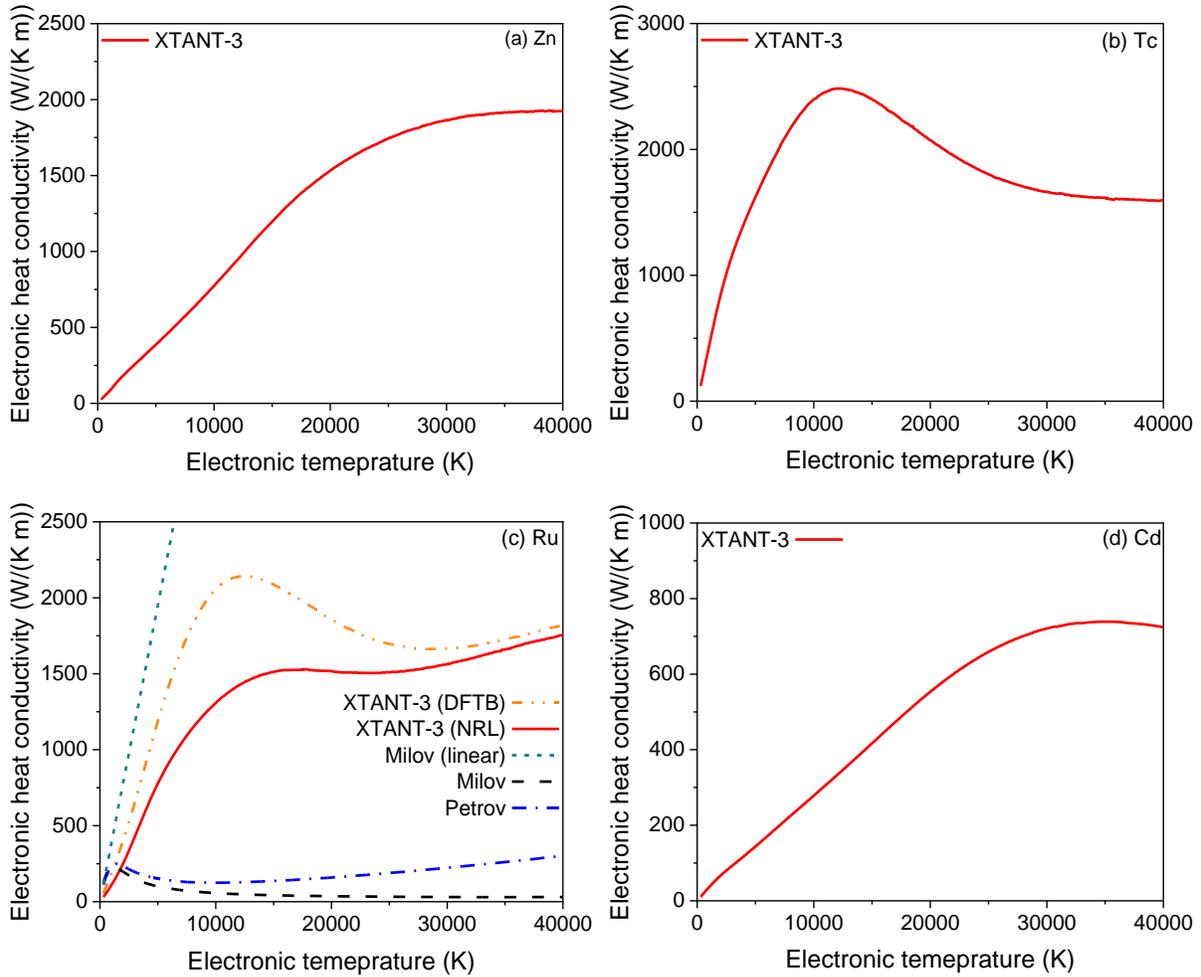

Figure 7. Electronic heat conductivity as a function of the electronic temperature. (a) in Zinc. (b) in Technetium. (c) in Ruthenium, a comparison of the XTANT-3 calculated conductivity (with NRL and DFTB) compared with those by Milov *et al.* (low-temperature linear approximation and a full dependence, Eq.(14)) [65]; and Petrov *et al.*[23]. (d) in Cadmium.

Figure 8 shows the electronic heat conductivity in hafnium, rhenium, and osmium. No data for comparison were found for these materials.

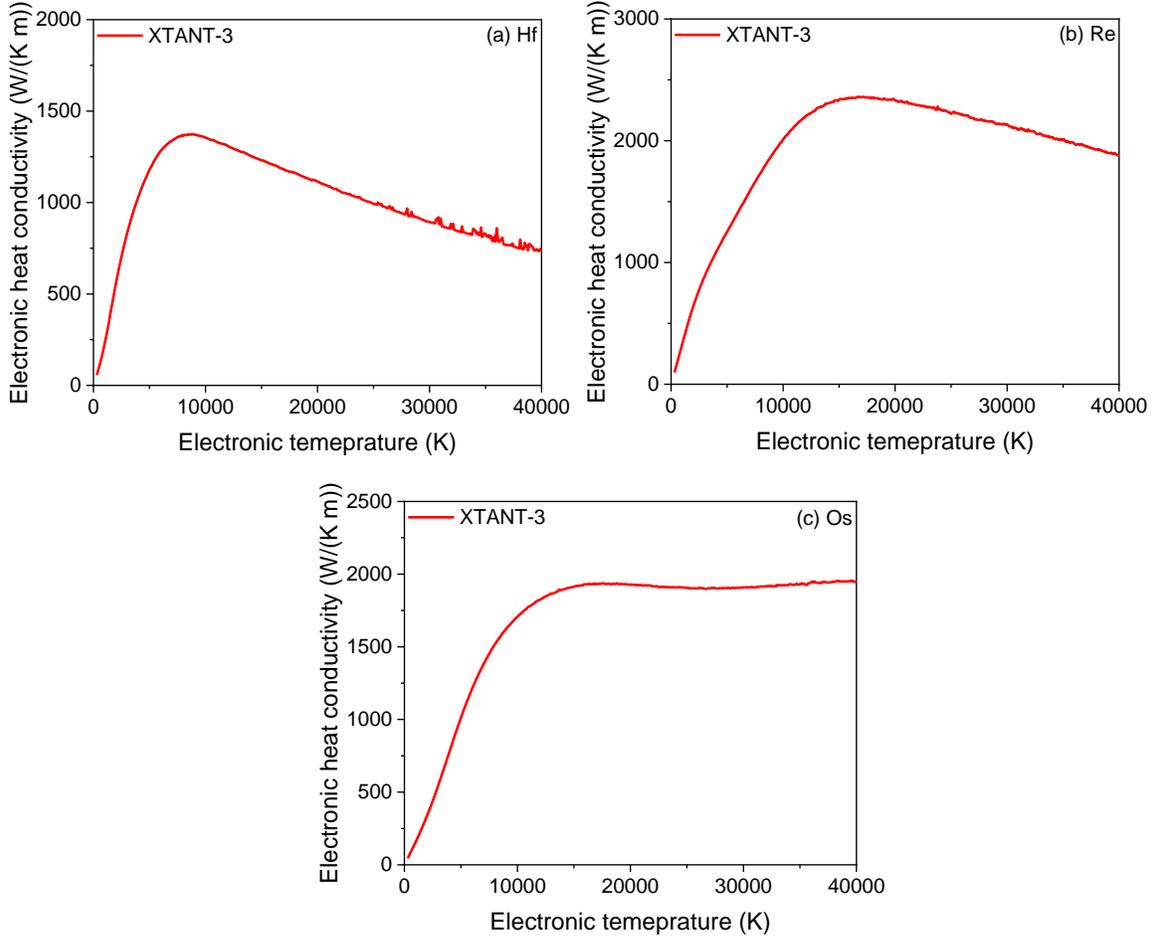

Figure 8. Electronic heat conductivity as a function of the electronic temperature. (a) in Hafnium. (b) in Rhenium. (c) in Osmium.

## 4. bcc metals

The following bcc metals are studied in this section: V, Cr, Fe, Nb, Mo, Ba, Ta, and W. We used 128 (4x4x4 unit cells) or 250 atoms (5x5x5 unit cells) in the supercell with NRL parameterization [34].

Figure 9 shows the results of calculations of the electronic heat conductivity in vanadium, chromium, iron, and niobium. The calculated electronic heat conductivity in chromium is compared with the model from Ref.[66], in which the authors assumed that the electronic heat conductivity in chromium is 0.6 of that in gold. In contrast, XTANT-3 calculations predict that the conductivity in chromium at high electronic temperatures is higher.

XTANT-3 calculated conductivity in iron is compared with those from Petrov *et al*.[44], showing a large discrepancy. It may be an effect of electron-electron contribution, although Ref.[44] does not show partial conductivities. Future experimental validation is required.

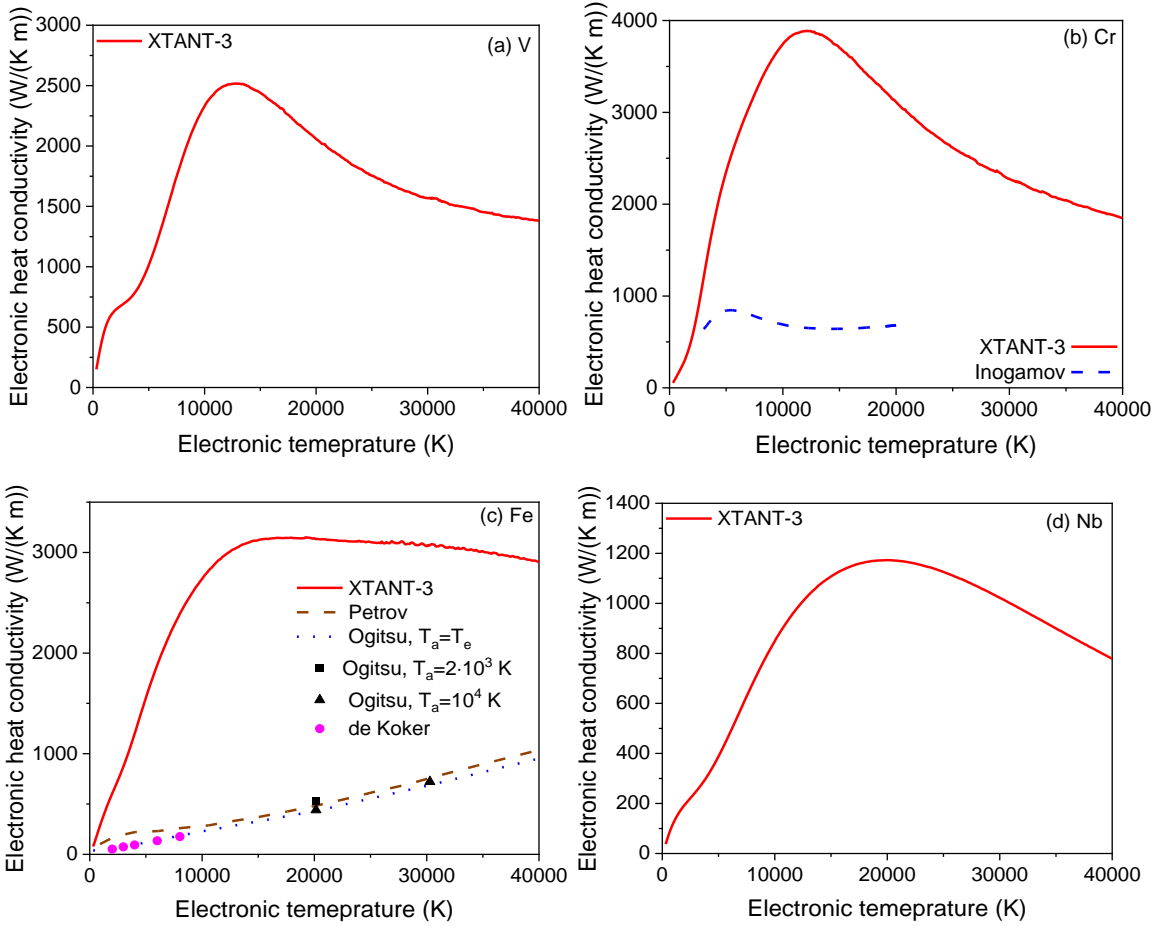

Figure 9. Electronic heat conductivity as a function of the electronic temperature. (a) in Vanadium. (b) in Chromium, XTANT-3 calculations vs. that from Inogamov *et al.*[66]. (c) in Iron, XTANT-3 calculated vs. those from Petrov *et al.* [44], Ogitsu *et al.*[67] and de Koker *et al.* [68]. (d) in Niobium.

Figure 10 demonstrates calculated electronic thermal conductivity in molybdenum, barium, tantalum, and tungsten. Molybdenum results, shown in Figure 10a, are compared to those by Olbrich *et al.*[51]. The initial rise of the curves is similar, but Olbrich *et al.*'s result continues to rise after $T_e$~10,000 K to twice as large values in the peak.

For Tantalum, our calculated conductivity is compared to that of Petrov *et al.* [44]. The XTANT-3 data are significantly larger than the semiempirical estimates from [44].

In Tungsten, the results are compared to the low-temperature estimation by Wang and Zhao for the case of $T_e=T_a$ reported in [69]. The same as in other cases discussed above, such an equilibrium scenario shows much smaller changes in the electronic conductivity, in comparison with a nonequilibrium case $T_e>T_a$, simulated with XTANT-3. As pointed out in Ref.[69], the low-temperature (near-room temperature) heat conductivity of materials is sensitive to many parameters, such as the material composition, presence of impurities, defects and dislocations, polycrystallinity, grain size, etc. Thus, low-temperature data are expected to deviate among various models as well as experiments.

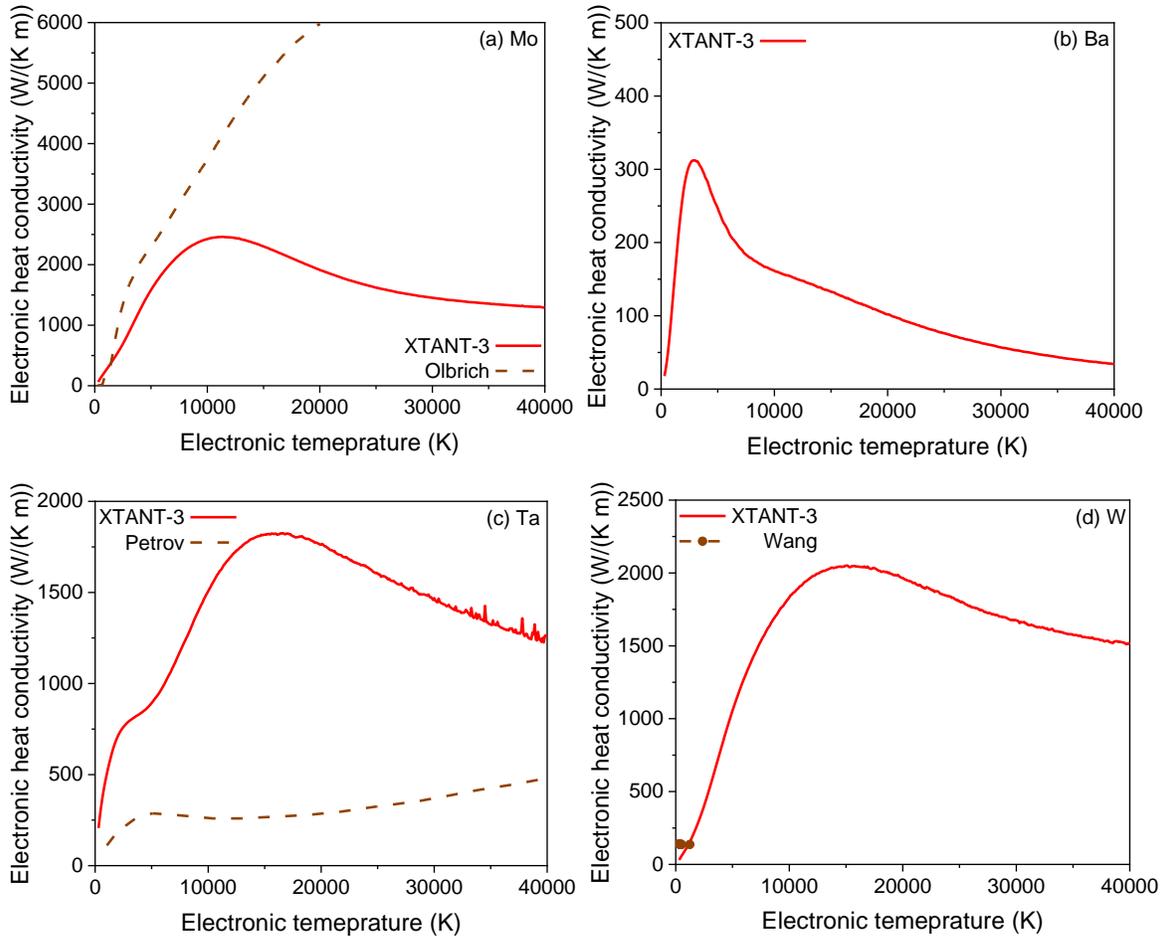

Figure 10. Electronic heat conductivity as a function of the electronic temperature. (a) In molybdenum, XTANT-3 calculated, compared to that by Olbrich *et al.*[51]. (b) in Barium. (c) in Tantalum, XTANT-3 calculated compared to those from Petrov *et al.* [44]. (d) in Tungsten, XTANT-3 calculated compared to low-temperature estimation by Wang and Zhao[69].

## 5. Other metals

A diamond cubic lattice metal Sn (216 atoms in the supercell); specific cases of Ga (144 atoms), In (144 atoms), Mn (232 atoms), Te (192 atoms), Se (192 atoms); and semimetal graphite (192 atoms) were modeled with NRL tight binding parameters [34]. The results for all of them are shown in Figure 11. Among those, the electronic heat conductivity reaches the highest values in graphite at high electronic temperatures of Te~30,000 K.

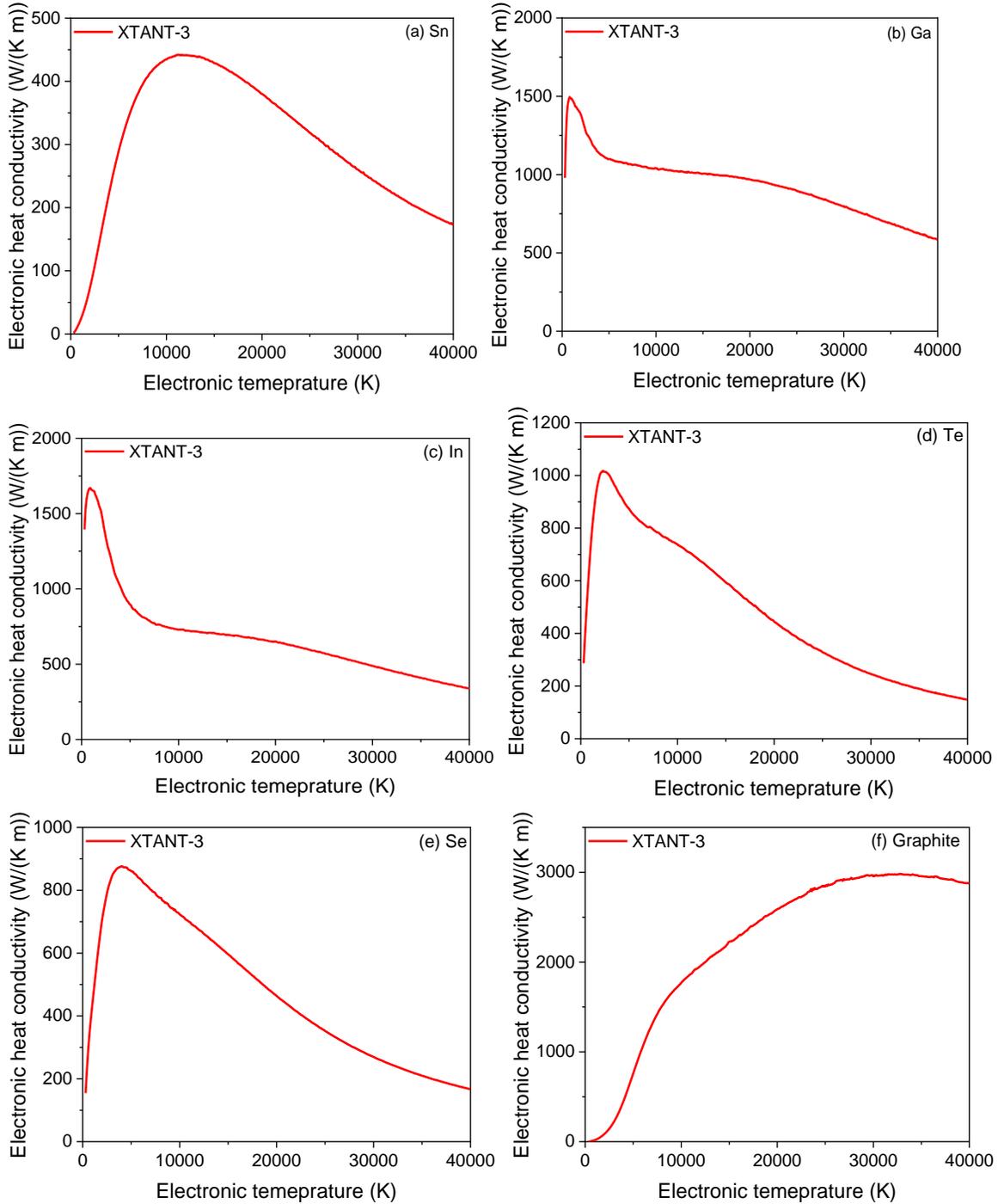

Figure 11. Electronic heat conductivity as a function of the electronic temperature. (a) in Tin. (b) Gallium. (c) in Indium. (d) in Tellurium. (e) in Selenium. (f) in Graphite.

## 6. Group IV semiconductors

For Si and Ge, the NRL tight-binding parameterization is used with 216 atoms in the simulation box (and a DFTB parameterization for Si, as discussed above) [34,70]; for SiC (in the hexagonal

P6$_3$mc state [71]), the matsci-0-3 DFTB parameterization is used with 192 atoms in the simulation box [35].

Figure 12 shows the electronic thermal conductivity in silicon, germanium, and SiC compound. In solid silicon (Figure 12a), we compare our results with the available experimental data from Glassbrenner and Slack [72], who extracted the electronic heat conductivity from the electronic conductivity data assuming the Widerman-Franz law. The results of DFTB-based XTANT-3 simulations at low electronic temperatures are reasonably close to the experimental data, validating our model. Note that the NRL-based calculations produce very similar functional dependence, but overestimate the experimental data by the absolute value at low electronic temperatures.

We also show a comparison with other available calculations: the ab initio-based Boltzmann model by Gu *et al*.[73]; the molecular dynamics-based Kubo-Greenwood calculations on the liquid Si by Migdal et al. from Ref. [51]; and the semiclassical model by Koroleva *et al*.[74]. At low electron temperatures, the XTANT-3 calculations with DFTB parameterization agree very well with the calculations by Gu *et al*. and with the results of the modeling by Koroleva *et al*.[74] up to the melting temperature of Si, see the inset in Figure 12a. At electron temperatures above the melting temperatures, the absolute values are comparable in all the results, but the rise of the conductivity with the electronic temperature is much faster in our calculations on solid Si than in the melted one (since data in the literature were obtained for $T_e=T_a$).

The data for germanium shown in Figure 12b compares the electronic heat conductivity calculated with NRL-based parameterization with the experimental data from Glassbrenner and Slack [72]. We note again that the functional dependence is reproduced correctly, but it overestimates the experimental curve by the absolute value. Recalling the results on silicon (DFTB vs NRL parameterizations), we may expect the high-electronic temperature results to be more reliable, but experimental validation is required. We also note that the experimental data are for the equilibrium case ($T_e=T_a$), which may also affect the comparison.

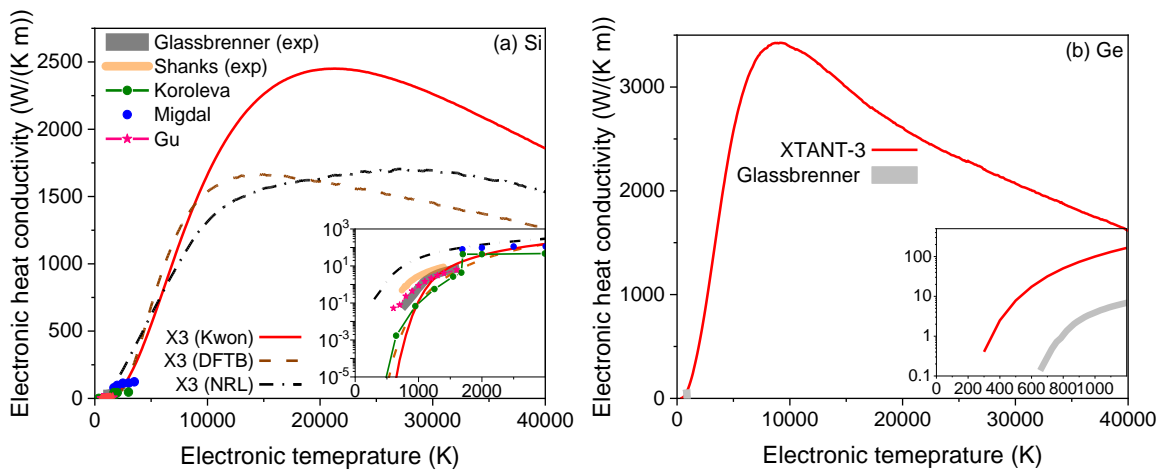

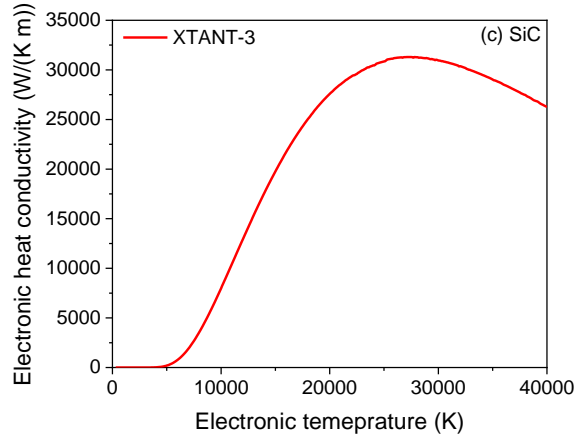

Figure 12. Electronic heat conductivity as a function of the electronic temperature. (a) in Si, XTANT-3 calculations (marked as "X3" in the legend, with three different TB parameterizations) compared with the experimental data by Glassbrenner and Slack [72]; and Shanks *et al.* [75] (using Eq.(4)); theoretical calculations by Gu *et al*. [73]; results of the model of liquid silicon by Koroleva *et al.*[74], and by Migdal *et al.*[76] (The inset zooms onto the low-temperature region, note the log scale). (b) in Ge, XTANT-3 calculations vs. the low-temperature experimental data by Glassbrenner and Slack[72]. (c) in SiC.

## 7. Group III-V semiconductors

For group III-V semiconductors (AlAs, AlP, GaAs, GaP, and GaSb) modeling, Molteni *et al.*'s TB parameterization was used [77], 216 atoms in the supercell in the zinc blende structure in each case. Except for GaSb, the other materials in this group show similar behavior, reaching comparable maximum values at electronic temperatures of $T_e$~30,000 K.

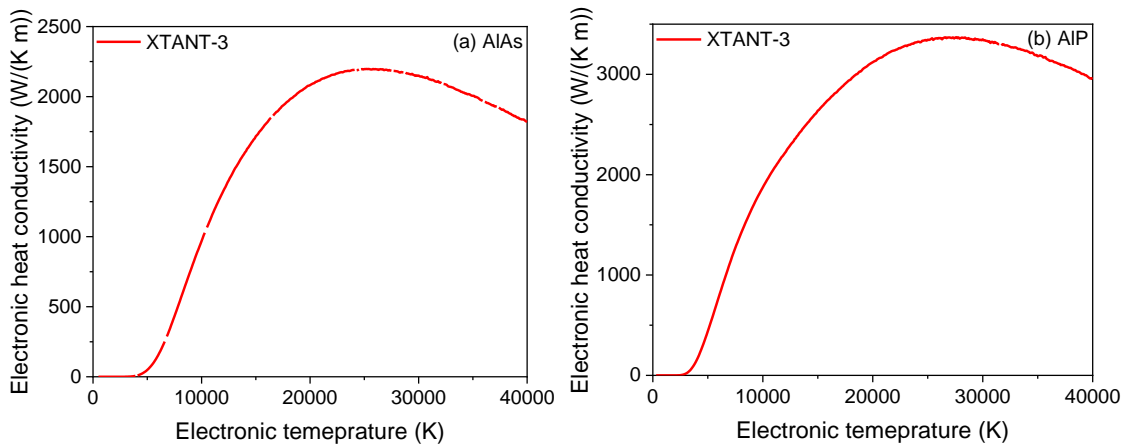

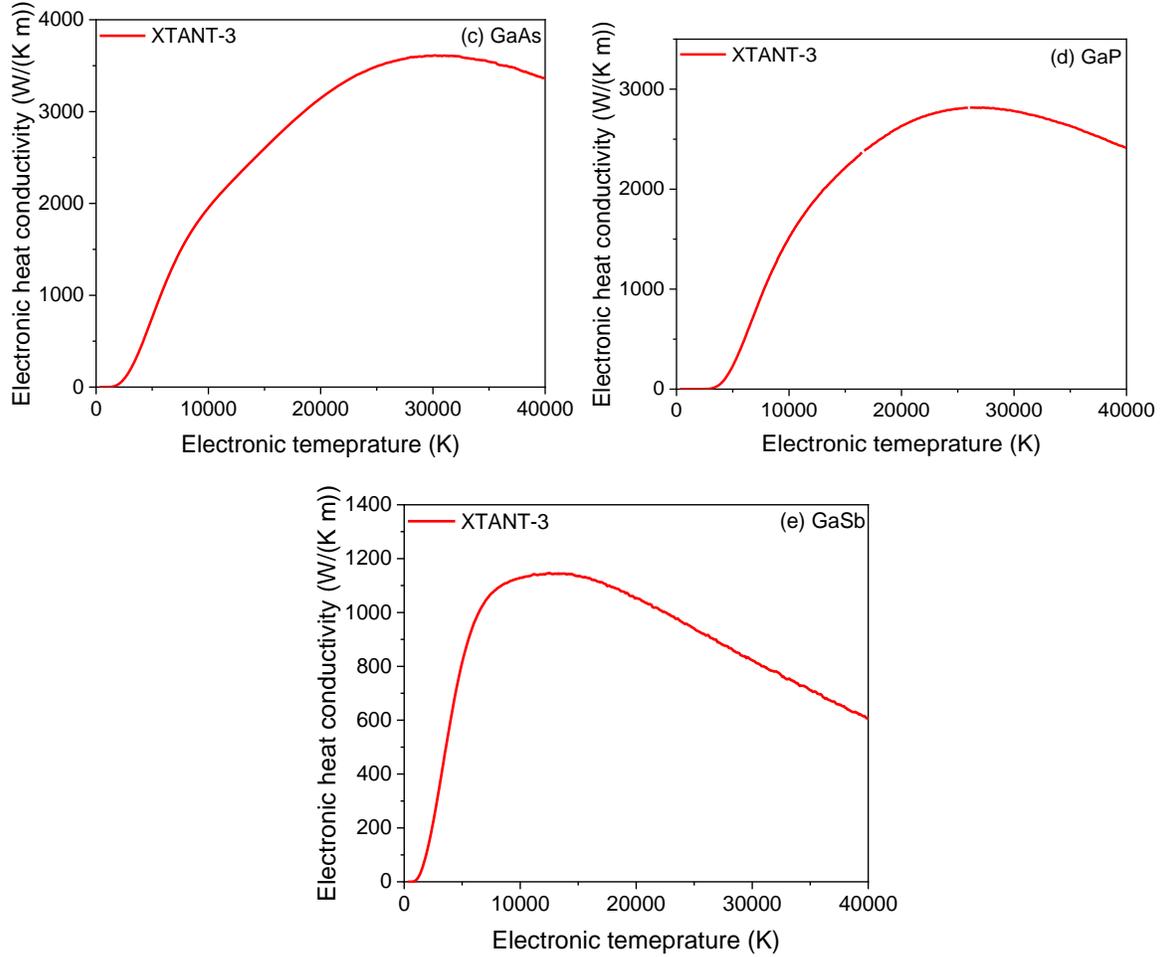

Figure 13. Electronic heat conductivity as a function of the electronic temperature. (a) in AlAs. (b) AlP. (c) in GaAs. (d) in GaP. (e) in GaSb.

## 8. Oxide semiconductors

For modeling of oxides semiconductors $Cu_2O$ (384 atoms in the simulation box in cubic Pn3m structure[71]) and $TiO_2$ (216 atoms in the simulation box in rutile structure), matsci-0-3 DFTB parameterization was used [35]; and znorg-0-1 DFTB parameterization for ZnO (384 atoms in hexagonal P6$_3$mc structure [71]) was simulated [78]. Oxides show very high peak electronic heat conductivities at temperatures $T_e$~30,000 K.

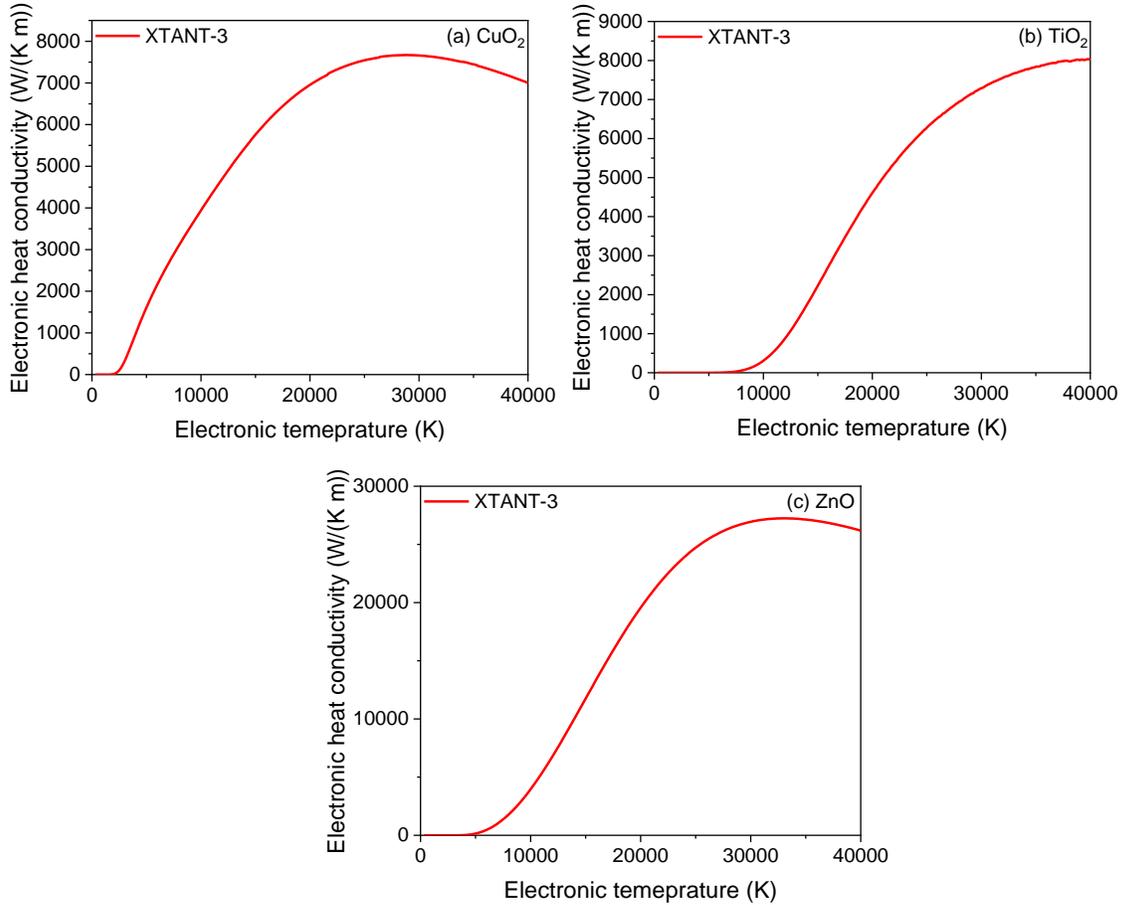

Figure 14. Electronic heat conductivity as a function of the electronic temperature. (a) in $Cu_2O$. (b) $TiO_2$. (c) in ZnO.

## 9. Other semiconductors

Three other types of semiconductors were also studied: $B_4C$ (270 atoms in trigonal R-3m structure [71]) using matsci-0-3 DFTB parameterization [35]; group II-VI semiconductor ZnS, modeled with znorg-0-1 DFTB parameterization (252 atoms in trigonal $P3m_1$ structure [71]) [78]; and layered $PbI_2$ (consisting of 192 atoms in hexagonal $P6_3mc$ structure [71]) with the DFTB-based parameterization from Ref. [79] (with added ZBL-short-range repulsive potential similar to the method described in Ref. [80]). $B_4C$ shows a nearly linear dependence of the electronic thermal conductivity with the electronic temperature in the studied range, reaching values comparable to those in oxides. The other two materials reach values smaller than those in other semiconductors.

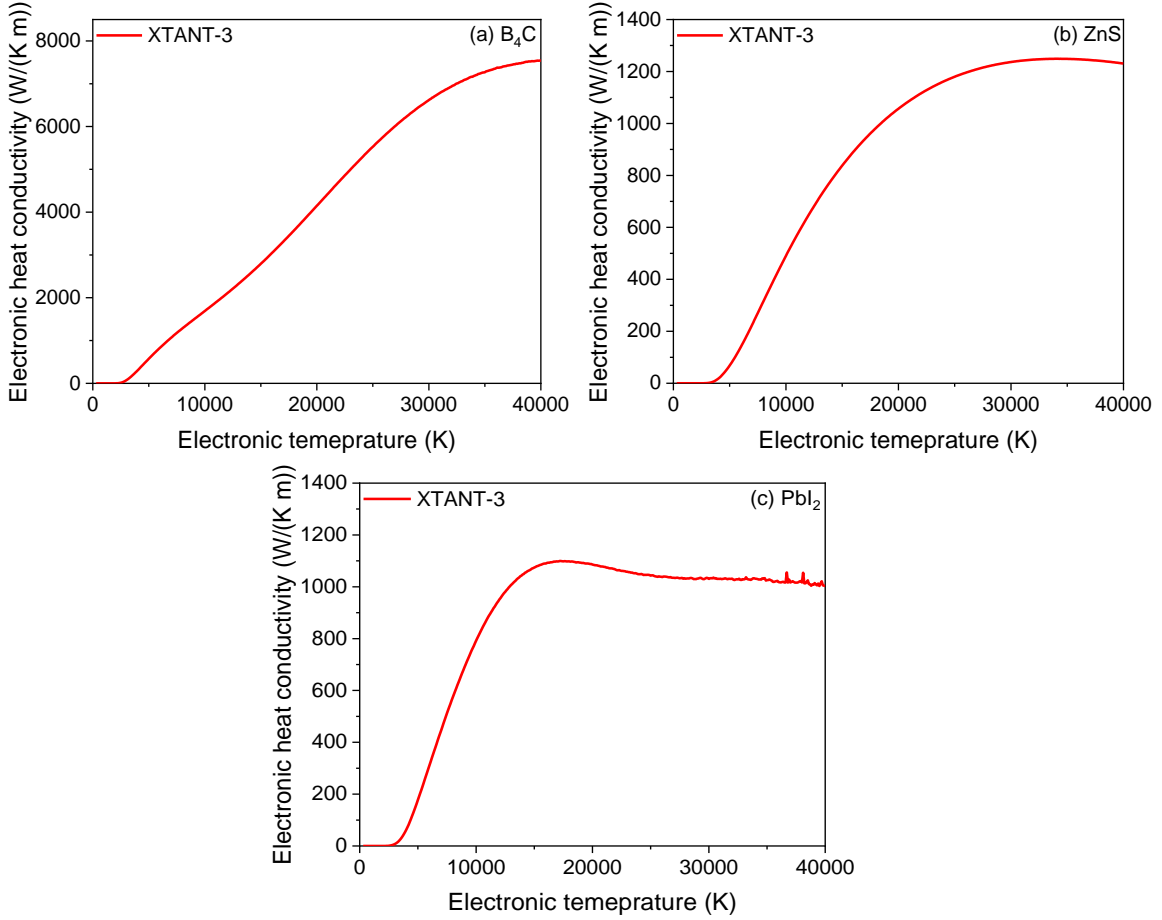

Figure 15. Electronic heat conductivity as a function of the electronic temperature. (a) in $B_4C$. (b) ZnS. (c) in $PbI_2$.

## IV. Discussion

Having a large dataset of the electronic heat conductivities amassed here, we can draw a few conclusions on the trends.

First, we see that, typically, semiconductors exhibit larger electronic heat conductivities than metals at high electronic temperatures. This counterintuitive behavior may be explained by the fact that at elevated electronic temperatures, the electronic heat conductivity is limited by the electron-electron scattering contribution. Indeed, in metals, electron-electron scattering is more significant due to the absence of the bandgap, thus, a higher frequency of scattering events lowers the conductivity.

Second, we can analyze the dependence of the electronic heat conductivity in elemental metals across the Periodic Table, see Figure 16. For this figure, we chose the values at the electronic temperature of $T_e$=10,000 K, similar to our previous analysis of the electron-phonon coupling in Ref.[17]. We see a similar trend of a decrease in the values with an increase in the electronic temperature on average, modulated by the peaks associated with the d-band metals. These trends seem to be mainly defined by the electron-electron contribution (Figure 16, middle panel). Those may be associated with the trends in the normal material densities having similar peaks (see discussion in Ref. [17]).

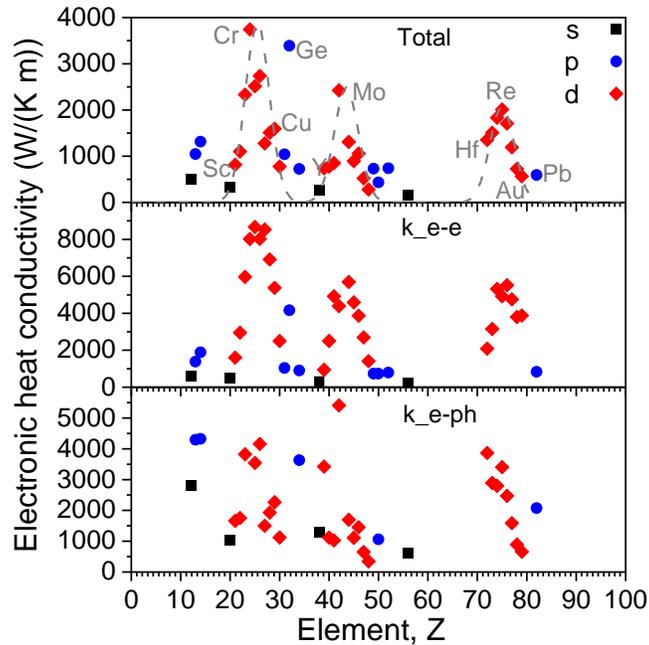

Figure 16. Electronic heat conductivity at $T_e$=10,000 K as a function of the atomic number in the periodic table for elemental materials: total (top panel); electron-electron contribution (middle panel); and electron-phonon term (bottom panel). Black squares mark materials with s-electrons in the outer orbital, blue circles mark p-electron materials, and red diamonds are used for d-electron materials.

## V. Conclusions

This work presented a model for calculations of the electronic heat conductivity at elevated electronic temperatures, relevant for laser-irradiation scenarios. The model is based on the tight-binding approach for evaluation of the electron-phonon contribution, and the linear response theory for electron-electron scattering contribution. It was implemented in XTANT-3 hybrid code and applied to various materials: metals across the Periodic Table, and a few elemental and compound semiconductors of various types.

Comparison with the experimental data available at low temperatures showed a reasonable agreement. The differences among various calculations reported in the literature are large, especially at high electronic temperatures, where in some cases the various models predict quantitatively different behavior. Thus, we envision that this work will motivate experimental validation of the predictions, resolving the controversies arisen due to the application of different models.

## Data availability statement

All the tables with the calculated electronic heat conductivities, reported here, are freely available online at https://github.com/N-Medvedev/XTANT-3_coupling_data, together with the electronic heat capacities and electron-phonon coupling parameters.

## Acknowledgments

Computational resources were supplied by the project "e-Infrastruktura CZ" (e-INFRA LM2018140) provided within the program Projects of Large Research, Development, and Innovations Infrastructures. NM gratefully acknowledges financial support from the Czech Ministry of Education, Youth, and Sports (grants No. LTT17015, LM2023068, and No. EF16_013/0001552).

[10] C.-W. Jiang, X. Zhou, Z. Lin, R.-H. Xie, F.-L. Li, R.E. Allen, Electronic and Structural Response of Nanomaterials to Ultrafast and Ultraintense Laser Pulses, J. Nanosci. Nanotechnol. 14 (2014) 1549–1562. https://doi.org/10.1166/jnn.2014.8756.

[11] F. Akhmetov, N. Medvedev, I. Makhotkin, M. Ackermann, I. Milov, Effect of Atomic-Temperature Dependence of the Electron–Phonon Coupling in Two-Temperature Model, Materials (Basel). 15 (2022). https://doi.org/10.3390/ma15155193.

[12] A. Caro, M. Victoria, Ion-electron interaction in molecular-dynamics cascades, Phys. Rev. A. 40 (1989) 2287–2291. https://doi.org/10.1103/PhysRevA.40.2287.

[13] D. Ivanov, L. Zhigilei, Combined atomistic-continuum modeling of short-pulse laser melting and disintegration of metal films, Phys. Rev. B. 68 (2003) 064114. https://doi.org/10.1103/PhysRevB.68.064114.

[14] R. Darkins, D.M. Duffy, Modelling radiation effects in solids with two-temperature molecular dynamics, Comput. Mater. Sci. 147 (2018) 145–153. https://doi.org/10.1016/j.commatsci.2018.02.006.

[15] Z. Lin, L. Zhigilei, V. Celli, Electron-phonon coupling and electron heat capacity of metals under conditions of strong electron-phonon nonequilibrium, Phys. Rev. B. 77 (2008) 075133. https://doi.org/10.1103/PhysRevB.77.075133.

[16] N. Medvedev, I. Milov, B. Ziaja, Structural stability and electron-phonon coupling in two-dimensional carbon allotropes at high electronic and atomic temperatures, Carbon Trends. 5 (2021) 100121. https://doi.org/10.1016/J.CARTRE.2021.100121.

[17] N. Medvedev, I. Milov, Electron-phonon coupling in metals at high electronic temperatures, Phys. Rev. B. 102 (2020) 064302. https://doi.org/10.1103/PhysRevB.102.064302.

[18] N. Medvedev, I. Milov, Contribution of inter- and intraband transitions into electron–phonon coupling in metals, Eur. Phys. J. D. 75 (2021) 1–6. https://doi.org/10.1140/EPJD/S10053-021-00200-W.

[19] N. Medvedev, Electron-phonon coupling in semiconductors at high electronic temperatures, Phys. Rev. B. 108 (2023) 144305. https://doi.org/10.1103/PhysRevB.108.144305.

[20] N. Medvedev, XTANT-3, (2023). https://doi.org/10.5281/zenodo.8392569.

[21] N. Medvedev, XTANT-3: X-ray-induced Thermal And Nonthermal Transitions in matter: theory, numerical details, user manual, Http://Arxiv.Org/Abs/2307.03953. (2023). https://arxiv.org/abs/2307.03953v1 (accessed July 11, 2023).

[22] N. Medvedev, V. Tkachenko, V. Lipp, Z. Li, B. Ziaja, Various damage mechanisms in carbon and silicon materials under femtosecond x-ray irradiation, 4open. 1 (2018) 3. https://doi.org/10.1051/fopen/2018003.

[23] Y. Petrov, K. Migdal, N. Inogamov, V. Khokhlov, D. Ilnitsky, I. Milov, N. Medvedev, V. Lipp, V. Zhakhovsky, Ruthenium under ultrafast laser excitation: Model and dataset for equation of state, conductivity, and electron-ion coupling, Data Br. 28 (2020) 104980. https://doi.org/10.1016/j.dib.2019.104980.

[24] V. Recoules, J.P. Crocombette, Ab initio determination of electrical and thermal
26